\documentclass[ preprint,3p]{elsarticle}
\usepackage{amsmath,amssymb,amsfonts,csquotes}
\usepackage{graphicx}
\usepackage{xcolor}
\usepackage{float}

\usepackage{hyperref}
\hypersetup{
colorlinks=true,
urlcolor= blue,
citecolor=blue,
linkcolor= blue,}

\usepackage{bm}

\usepackage{natbib}

\begin{document}
\title{Similarities among top one day batters: physics-based quantification}
\author[1]{Dipak Patra\corref{cor1}%
\fnref{fn1}}
\ead{dipak@rri.res.in}
\cortext[cor1]{Corresponding author}

\affiliation[1]{organization={Soft Condensed Matter Group, Raman Research Institute}, 
                 addressline={C. V. Raman Avenue, Sadashivanagar},
                 city={Bangalore-560080},
                 country={India}}

\begin{abstract}
 Assessment of the performance of a player in any sport is very much needed to determine the ranking of players and make a solid team with the best players. Besides these, fans, journalists, sports persons, and sports councils often analyse the performances of current and retired players to identify the best players of all time. Here, we study the performance of all-time top batters in one-day cricket using physics-based statistical methods. The batters are selected in this study who possess either higher total runs or a high number of centuries. It is found that the total runs increases linearly with the innings number at the later stage of the batter carrier, and the runs rate estimated from the linear regression analysis also increases linearly with the average runs. The probability of non-scoring innings is found to be a negligibly small number (i.e., $\leq 0.1$) for each batter. Furthermore, based on innings-wise runs, we have computed the six-dimensional probability distribution vector for each player.  
Two components of the probability distribution vector vary linearly with average runs. The component representing the probability of scoring runs less than $50$ linearly decreases with the average runs. In contrast, the probability of scoring runs greater than or equal to $100$ and less than $150$ linearly increases with the average runs. We have also estimated the entropy to assess the diversity of a player. Interestingly, the entropy varies linearly with the average runs, giving rise to two clusters corresponding to the old and recent players. Furthermore, the angle between two probability vectors is calculated for each pair of players to measure the similarities among the players. It is found that some of the players are almost identical to each other.
\end{abstract}
 \begin{keyword}
Sports data analysis \sep cricket \sep probability distribution \sep entropy  
\end{keyword}

\date{March 18, 2024}
\maketitle

\section{Introduction}
The game of cricket is the world's second most popular sport, after soccer, with over a billion fans \citep{Fans_2018}. Since its discovery, the game has continuously evolved into three formats: ``Test'', ``One Day International'' (ODI), and ``Twenty20'' (T20), and became popular across the countries. Countries like India, Australia, South Africa, Pakistan, New Zealand, England, Sri Lanka, Bangladesh, New Zealand, and the West Indies host most international cricket matches. 
In ODI cricket, a match consisting of two innings is played between two teams with 11 players, including batters, bowlers, all-rounders, and wicket-keeper, and it is started by tossing a coin. The captain of a team who wins the coin toss decides to either bat or bowl for the team's action in the first innings of the match. The team that performs the batting in the first innings tries to set up a target for the bowling team by scoring as many runs as possible. In the second innings, the other team goes for batting and tries to exceed the target score. A team performs a maximum of 50 bowling overs in an innings. In an over, a bowler consecutively throws the cricket ball six times towards the batters. However, an innings is terminated before the quota of 50 overs when the ten players of the batting team are dismissed. The team scoring the maximum number of runs in an innings is the winner of the match.

\begin{table}\label{tab:players}
\caption{List of selected top ODI batters for the study. The current active batters are denoted by *. }
\begin{tabular}{p{0.7in} p{1.4in} p{0.75in} p{0.6in} p{0.5in} p{0.5in} p{0.8in}}
Serial No & Batter Name & Span & Innings & Total Runs & Hundred & Average runs \\
\hline
1 &SR Tendulkar & 1989--2012& 452 & 18426 & 49 & 40.76 \\
2 & KC Sangakkara & 2000--2015& 380  & 14234 & 25 & 37.46\\
3 & V Kohli* & 2008--2023 & 280  & 13848 & 50 & 49.46\\
4 &RT Ponting  & 1995--2012 & 365 &  13704 & 30 &37.54\\
5 & ST Jayasuriya  &1989--2011 & 433  & 13430 & 28 &31.02\\
6 & DPMD Jayawardene  & 1998--2015& 418 & 12650  &19 &30.26\\
7& Inzamam-ul-Haq & 1991--2007 & 350 & 11739 & 10 &33.54\\
8 & JH Kallis  & 1996--2014 &  314 & 11579 & 17 &36.88\\
9 &SC Ganguly  &1992--2007 & 300  &  11363 &22 &37.88\\
10 & R Dravid & 1996--2011 & 318 &  10889  & 12&34.24\\
11 &  Ms Dhoni   &2004--2019 & 297 & 10773 & 10&36.27\\
12 &  RG Sharma* &2007--2023 & 254  & 10709 &31 &42.16\\
13  &CH Gayle  &1999--2019 & 294  & 10480 &25 &35.65\\
14 & BC Lara  &1990--2007 &  289 & 10405  &19 &36.00\\
15 &TM Dilshan &1999--2016 & 303  & 10290 &22 &33.96\\ 
16 & AB DeVilliers  & 2005--2018 & 218 & 9577 &25 &43.93\\
17 & Saeed Anwar &1989--2003 & 244 & 8824  & 20 &36.16\\
18 & LRPL Taylor  &2006--2022 & 220 &  8607 & 21&39.12\\
19 & HM Amla   &2008--2019 & 178 & 8113 &27 &45.58\\
20 & HH Gibbs  &1996--2010 & 240 & 8094 &21 &33.72\\
21 & DA Warner* &2009--2023 & 159 & 6932 & 22 &43.6\\
22 &Q de Kock* & 2013--2023 & 155 &  6770  & 21 &43.68\\
\hline
\end{tabular}
\end{table}

A large number of studies have been conducted to assess various fields in a match, including evolution and prediction of score, contribution of players, and performance of a team \citep{Ram_2022, Petersen_2008, Bhattacharjee_2014, Shanto_2019, Perera_2016}. In addition, the assessment of the performances of players throughout their careers has been studied over the past. The findings of the assessment studies are crucially needed for the team selection, ranking of players, and determining all-time best players \citep{Kimber_1993, Staden_2009, Damodaran_2006, Beaudoin_2003}. Because of the implication, these studies are greatly embraced by various sectors of societies, including team management/selectors, coaches, sports councils, and sports industries. Particularly, in a debate, the comparison (including differences and similarities) among batters are often drawn by the fans and journalists to validate their argument \citep{Narayanan_2020, Nicholas_2023}.
The batters are mainly evaluated using classical methods such as batting average or strike rate. However, these classical quantities are limited because of their biases \citep{Narayanan_2020_1}. Therefore, a wide variety of tools
such as principal component analysis \citep{Manage_2013, Gupta_2022}, network theory \citep{Mukherjee_2014} and machine learning \citep{Iyer_2009} have been employed for the evaluation of the batters. However, the quantity that can measure the diversity of a batter is not addressed in the earlier studies. In particular, diversity is vital to a player as it is directly associated with the ability of a player to score runs across different levels, such as fifties, hundreds, two-hundreds, etc. Furthermore, these studies lack the measurable quantity determining the similarities between the players. However, the study exploring the diversity and similarities of players can impact the selection of players for making a balanced team. Also, it is expected to be cost-effective by setting a reasonable bid price range for identical players. Thus, it can reduce the enormous price fluctuation during the Indian Premier League auction \citep{IPL_2024}.

It is well known that the game of cricket is unpredictable because of its stochastic nature. The outcomes of the games can be represented as the time series data.
Therefore, the physics-based system consisting of Brownian particles has been employed to study the dynamics of the score evolution in a match for all three formats of cricket games \citep{Ribeiro_2012}. Furthermore, the tools describing physical systems have been utilized to study various other fields such as stock market \citep{Ausloos_2002, Schmitt_2012, Gunduz_2017}, crowd dynamics \citep{Helbing_2000, Silverberg_2013}, disease spreading \citep{Forgacs_2023} and sports \citep{Ben-Naim_2007, Chacoma_2022} especially soccer \citep{Bittner_2007, Bittner_2009, Heuer_2010, Stock_2022}. Surprisingly, despite the profound implications across different fields, physics-based tools have not been significantly explored in cricket.

Here, we assess the performance of the top twenty-two ODI batters by employing physics-based tools. The growth dynamics of the total runs for all players are investigated. The probabilities of scoring runs across six different score levels are also computed to construct a six-dimensional probability distribution vector for each player. Some components are directly related to the number of fifties and hundreds achieved by a player. Furthermore, we assess the diversity of a player by computing the entropy from the probability distribution vectors. The angles between two probability distribution vectors corresponding to each pair of batters are also calculated to measure the similarities between the corresponding players. After determining the angle for each pair of players, a similarity matrix is constructed. To understand the similarity matrix, we also determine the positions of the players on a cross-section of a spherical surface based on their performances. 

\section{Methods}
 We have collected data indicating the performances of the batters in one-day cricket from the Cricinfo website \citep{Cricinfo}. In this study, the performances of two types of batters have been assessed depending on either total runs or total hundreds scored by the players. Therefore, we have selected the batters whose total runs or hundreds are greater than ten thousand or nineteen, respectively. The Cricinfo website provides that the statistics of twenty-two players up to the date 20/11/2023 satisfy these constraints. 
The total runs and centuries of each of these batters are indicated in table.~\ref{tab:players}.
We have assessed the data representing the innings number vs runs for each of the 22 players.
The matches where the batters did not bat have been excluded from this study. We have also assumed the not-out innings as the end of the innings for the player; therefore, the score of not-out innings has not been modified. 
 It should be noted that the innings number is analogous to the discrete-time interval in the time series data. Therefore, the innings numbers can be treated as dimensionless time units.
   We have calculated the total runs from the innings-wise runs and studied the variation of total runs as a function of time/innings number to determine the nature of growth dynamics for each batter.
 The average runs (AR) for each player is calculated based on the relation
 \begin{equation}
  AR = \sum^{N}_{k=1} \frac{R_{k}}{N}~,
 \end{equation}
 where $R_{k}$ is the runs scored by a batter at $k$-th innings, and $N$ is the total number of innings played by the batters. In table.~\ref{tab:players}, the average runs corresponding to the batters are also noted.
It is also assumed that a batter can score a maximum of 300 runs in an innings.
Innings-wise runs corresponding to each batter are distributed over six score levels. The first score level is associated with runs less than 50. The probability of scoring run in the first level is denoted by $p_1$. The probability of scoring runs greater than equal to 50 and less than 100 is accounted for by $p_2$. The probability $p_3$ connects with the third score level describing the scoring runs greater than equal to 100 and less than 150. The probability corresponding to the fourth score level relating the scoring run greater than equal to 150 and less than 200 is denoted by $p_4$. The fifth score level represents the scoring run greater than or equal to 200 and less than 250, and $p_5$ indicates the probability of the corresponding level. The last score level is associated with the scoring runs greater than 250, and its probability is noted as $p_6$.
The cell-wise probability distribution for each player is represented as a six-dimensional vector $P=(p_1, p_2, p_3, p_4, p_5, p_6)$. The component of the six-dimensional probability distribution vector is utilized to measure the diversity of the batters. The diversity can be assessed by determining the entropy.
 The entropy for each player is defined as
\begin{equation}
 S=-\sum_i^6 p_i log p_i ~.
\end{equation}
In physics, this entropy expression is known as Gibbs entropy, which is utilized to assess the thermodynamical properties of a system. On the other hand, it is known as Shanon entropy in information theory, statistics, machine learning, and image analysis. 
The six-dimensional probability distribution vectors of the batters are also used to measure the similarities among players.
The similarities between the two players are assessed by finding the angle between their unit probability distribution vectors.
The element of the similarity matrix $A$ is defined as
\begin{equation}
A_{ij}= cos^{-1} \hat{P}_i \cdot \hat{P}_j,
\end{equation}
where $P_i$ and $P_j$ are the probability distribution vectors for $i$-th and $j$-th batter respectively. A similar type of vector space model has also been used for finding out the identical documents and automatic indexing \citep{Salton_1975}.
 \begin{figure}[t]
\includegraphics[clip=true,width=\columnwidth]{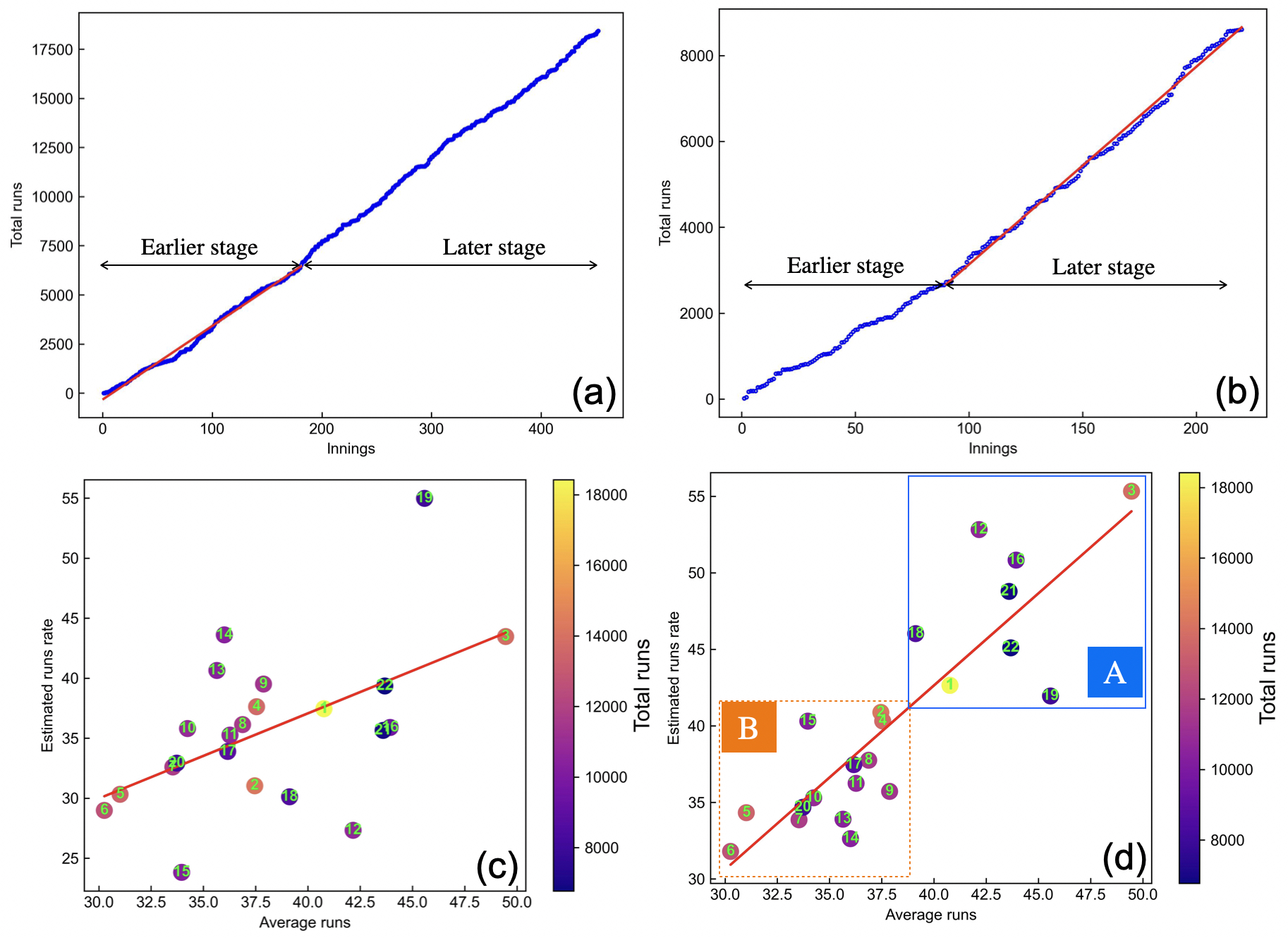}
   \caption{The variation of total runs as a function of innings number (time) for the batters (a) Tendulkar and (b) Taylor. The earlier and later career stages of the players are annotated in these graphs. The variation of estimated runs rate (velocity) corresponding to the earlier and later career stages of batters as a function of average runs shown in (c) and (d), respectively. 
   For convenience, the serial numbers of batters are annotated in the graphs. Two boxes representing groups marked as ``A" and ``B" are schematically drawn. Group ``A" consists of mostly recent or active players, whereas Group ``B" only contains old-retired players. The red lines are drawn based on the linear regression results. The color-coded bars display the total runs scored by the players during their career span.}
    \label{fig1}
\end{figure}

\section{Result and Discussion}
We have assessed the performances of top batters in one-day cricket by deploying various methods described above.
Figure~\ref{fig1}(a) and (b) represent the variation of total runs ($R$) with innings number ( time ) for Tendulkar and Taylor, respectively. A similar type of non-linear variation is obtained for other players. At lower values of time, the total run increases in a slightly non-linear fashion. This perhaps explains that the players are inexperienced and perform inconsistently in their earlier innings. As time passes, players become consistent. Therefore, the total run increases more or less linearly for higher values of time (i.e., $R \sim t^\alpha$), leading to the growth exponent $\alpha$ equal to 1. The growth dynamics of total runs do not follow diffusive processes wherein the growth exponent $\alpha$ equals 1/2 (i.e., $\sim \sqrt{t}$). Similar growth dynamics have been observed in other sports and scientific fields. For example,  the career trajectory of Major League Baseball and National Basketball Association players shows the growth dynamics with $\alpha=1$ \citep{Petersen_2012} whereas the career growth of scientists displays slightly higher values of growth exponent with $1 \leq \alpha \leq 1.5$ \citep{Petersen_2012, Sinatra_2016}.

To estimate the runs rate (velocity) for each player, we have performed the linear regression analysis against the total runs as a function of time. The two-run rates corresponding to the earlier and later career stages of a player are obtained from the linear regression analyses. The earlier career stage of a player is defined up to the $40\%$ of the total innings played by the player. The rest of the innings (i.e., $60\%$ of the total innings) are included in the later career stage of the player.
The velocities (runs rates) as a function of average runs are indicated in figure~\ref{fig1}(c) and (d) for both earlier and later career stages of the batters, respectively.
The velocity for the earlier stage varies non-linearly with the average runs. However, a linear fitting is performed, and the slope is obtained around 0.71. In this stage, it is difficult to classify the players into groups.
On the other hand, for the later career stage, the velocity increases linearly with the average runs, and it is fitted with a linear function, giving rise to the slope around 1.20. The higher slope in the later stage is expected because of the player's vast experiences and consistent performances.
In this case, the players can be divided into two groups based on the values of velocity and average runs. The group corresponding to the higher values of velocity and average runs is denoted by the ``A" class. In contrast, the group corresponding to the lower values of velocity and average runs is denoted by the `` B" class. Interestingly, group ``A" includes mostly recent or current batters, whereas group ``B" consists of only the earlier or retired batters. The points corresponding to the batters with serial numbers 1 (Sachin) and 18 (Taylor) can be considered the boundary points between these two groups.

It should be noted that the dynamics of an active Brownian particle can account for this linear growth of the total runs. At a minimal level, the Langevin equation describing the motion of the active Brownian particle in the over-damped limit is expressed as 
\begin{equation}\label{eq:brownian}
\gamma \frac{dR}{dt} = v + \eta,
\end{equation}
where $v$ is the active (drift) velocity of the particle due to the active force term, and the noise term $\eta$ acts as a random force on the particle. $\gamma$ represents the viscosity coefficient due to the fluid medium.
The total runs $R$ represents the instantaneous position of the Brownian particle in one dimension. The mean of the noise term is assumed to be zero, and its standard deviation is denoted by $\sigma$. Hence, the properties of the Gaussian noise $\eta$ are described as
\begin{eqnarray*}
< \eta > &=&0 \\
<\eta(t_1)\eta(t_2)>&=&\sigma^2 \delta(t_1 -t_2),
\end{eqnarray*}
where $\delta$ is the Kronecker delta function and the angular bracket $<>$ represents the ensemble average or averaging over the noise realization.
The average of the total runs $<R>$ can be expressed as
\begin{equation}
\frac{d<R>}{dt} =\frac{ v}{\gamma}.
\end{equation}
Thus, it increases on average linearly with time, accounting for the linear growth of the total runs of each player. 
It should be noted that this equation~\ref{eq:brownian} can be converted into another
type of Langevin equation describing a passive Brownian particle, such as 
\begin{equation}
\gamma \frac{dy}{dt}= \eta,
\end{equation}
where $y=R-(v/\gamma)t$ represents the position of the passive particle.
After doing some trivial algebra, one can 
obtain the variance of total runs as $\langle y^2 \rangle = \frac{\sigma^2}{\gamma^2} t$.
This simple model shows that the variance follows the normal diffusion process.
However, it has been found in the earlier study that the average score of a cricket innings shows linear growth with time (over). In contrast, the variance of the average score exhibits anomalous diffusion \citep{Ribeiro_2012}. Therefore, the variance of total runs can show
anomalous diffusion. In that case, our simple model needs to be extended to the general Langevin model, and the performances of a large number of batters need to be assessed. The study regarding anomalous diffusion deserves further investigation and is therefore aimed for future research. 

\begin{figure}[H]
\centering
   \includegraphics[height=10cm, width=14cm,keepaspectratio]{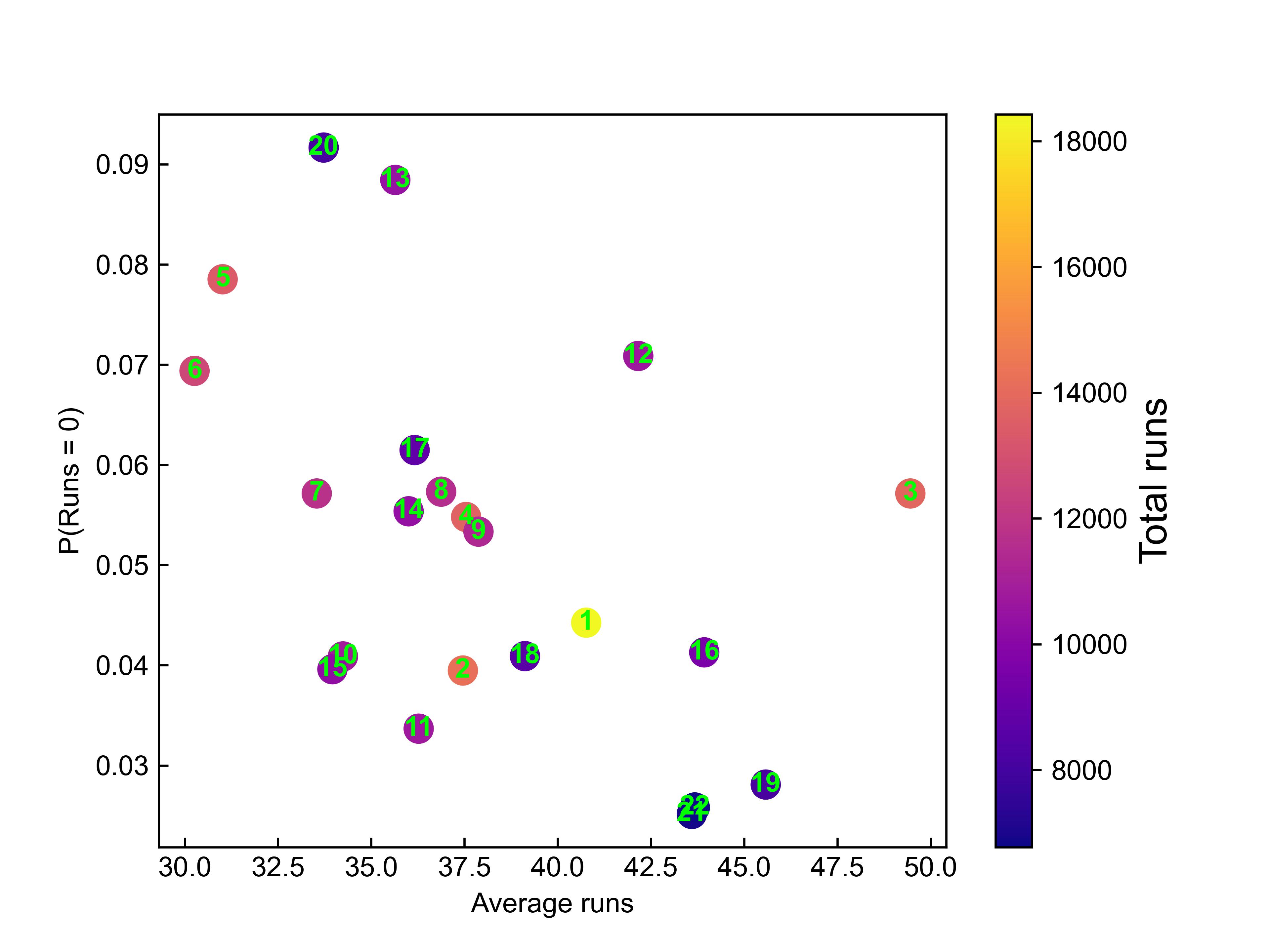}
   \caption{The variation of the probability of non-scoring innings as a function of average runs where annotated numbers representing serial numbers of the batters. The color-coded bar displays the total runs scored by the players during their career span.}
    \label{fig2}
\end{figure}

We have estimated the probability of non-scoring innings for each batter as shown in figure~\ref{fig2}. This probability is a crucial quantity to assess the quality of a batter. In principle, the lower value of it is appreciated by any cricket team. Figure~\ref{fig2} shows that the probability slightly decreases with increasing values of average runs. This probability is found to be minimum and maximum for the batters with serial numbers 21 (Warner) and 20 (Gibbs), respectively. 
Also, this probability is less than $0.1$ for each batter in our study. The lower value is expected for these top-class batters.

\begin{figure}[h]
\includegraphics[clip=true,width=\columnwidth]{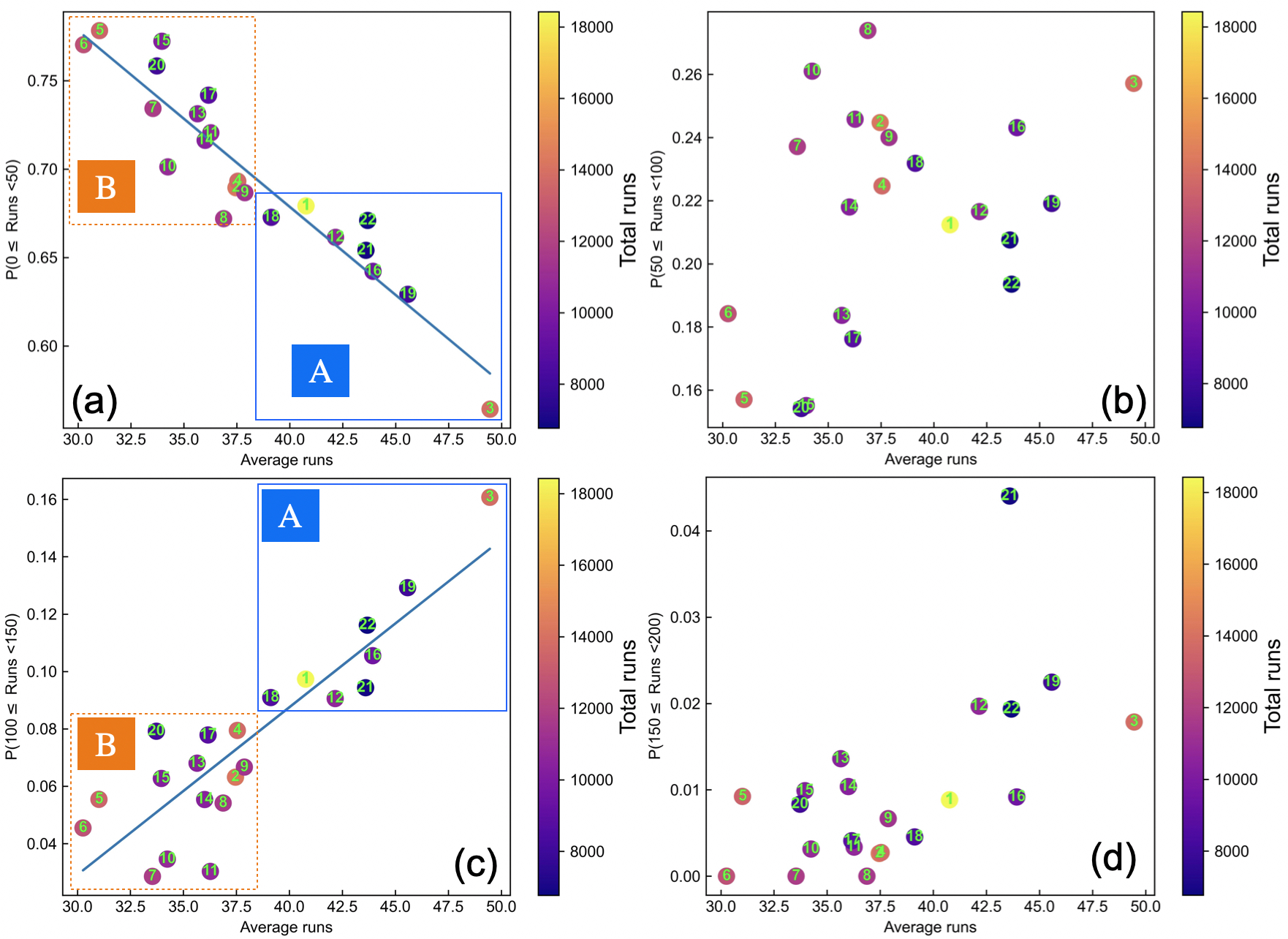}
   \caption{The variation of probability of scoring runs over different score levels as a function of average runs (a), (b), (c), and (d). For convenience, the serial numbers of batters are annotated. Two boxes representing groups marked as ``A" and ``B" are schematically drawn. Group ``A" consists of mostly recent or active players, whereas Group ``B" only contains retired players. The total runs scored by the players during their career span are represented by color-coded bars. The blue lines are drawn based on the linear regression results.}
    \label{fig3}
\end{figure}

We have also calculated the probability distribution of runs corresponding to the six score levels for each player. Figure~\ref{fig3}(a) represents the probability of scoring run in the first level ($p_1$) as a function of average runs. It is found that $p_1$ decreases with average runs, giving rise to the linear relationship. The slope estimated from the linear fit is of the order of $-0.0099$.
 It is expected that the higher value of average runs needs a large number of innings with higher values of scoring runs. 
Thus, the probability $p_1$ is minimum and maximum for the batters with serial numbers 3 (Kohli) and 5 (Jayasuriya), respectively. Players can be classified into two groups, ``A" and ``B", as discussed earlier. Figure~\ref{fig3}(b) represents the probability of scoring run in the second score level ($p_2$) as a function of average runs. The probability $p_2$ does not follow any particular trend and varies randomly with the average runs. A similar type of classification of the players into ``A" and ``B" groups is not warranted. The maximum of $p_2$ describes that the batter with serial number 8  (Kallis) has a maximum number of fifties in one-day cricket among the batters. In contrast, the batter with serial number 20 (Gibbs) has a minimum number of fifties, giving rise to a minimum value of $p_2$.
 In contrast to $p_1$ and $p_2$, the probability corresponding to the third score level $p_3$ increases linearly with the average runs as shown in figure~\ref{fig3}(c). The slope estimated from the linear fit is of the order of $0.0058$.
 This probability is found to be maximum for Kohli, which is expected as he possesses the highest number of hundreds in one-day cricket. One can easily divide the players into classes consistent with ``A" and ``B" groups. Figure~\ref{fig3}(d)
 indicates the variation of $p_4$ with runs average. The batter with serial number 21 (Warner) has scored in this level for the highest number of innings, giving rise to a maximum value of $p_4$.
  However, for all the players, $p_4$ is found to be less than $0.05$. The probabilities $p_5$ and $p_6$ are almost equal to zero for most of the players. Therefore, the scoring runs across these last two levels can be treated as an extremely rare event in a one-day cricket match.
 It should be noted that the probabilities corresponding to the score levels follow 
an empirical relation $p_1 \geq p_2 \geq p_3 \geq p_4 \geq p_5 \geq p_6$. This relation describes that scoring more runs in an innings is a daunting task.

\begin{figure}
\centering
   \includegraphics[height=8cm, width=14cm,keepaspectratio]{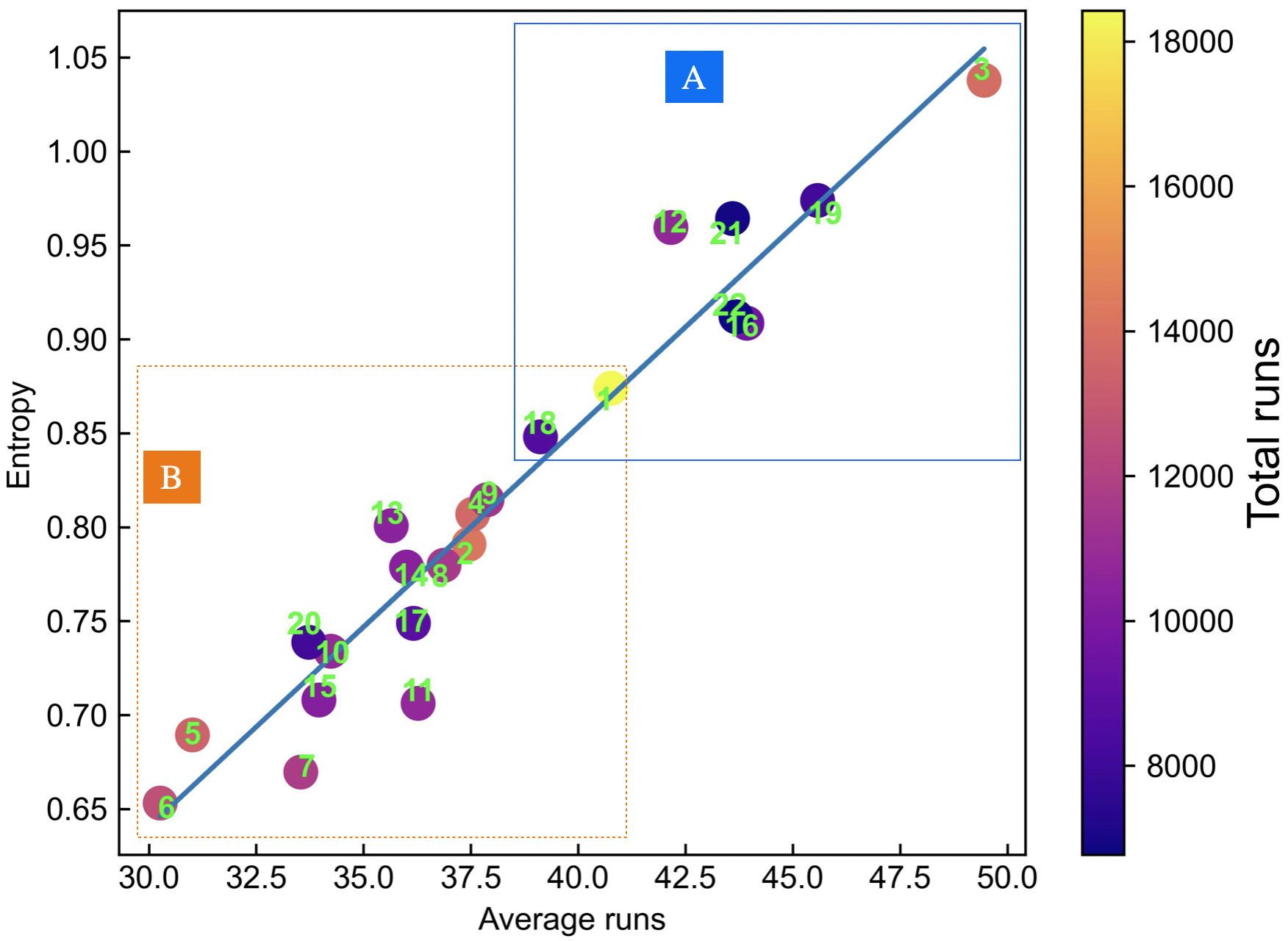}
   \caption{The variation of the entropy as a function of average runs. The annotated numbers indicate the serial numbers of batters. Two boxes representing two groups marked as ``A" and ``B" are schematically drawn. Group ``A" consists of mostly recent or active players, whereas Group ``B" only contains retired players. The total runs scored by the players during their career span are represented by a color-coded bar. The blue line is drawn based on the linear regression results.}
    \label{fig4}
\end{figure}

The probabilities corresponding to the six score levels determine the probability distribution vector for each batter.
Based on the probability distribution vector, we have calculated the entropy for each player. The entropy is a non-negative quantity measuring the value of disorder in a system. A perfectly order state corresponds to the value of entropy equal to zero.
 Figure~\ref{fig4} represents the variation of entropy as a function of average runs. It is found that the entropy varies linearly with the average runs, and the slope estimated from the linear fit is about $0.021$. 
It is generally known that the equilibrium state of a physical system corresponds to the maximum value of the entropy. So, the higher value of the entropy is associated with the ability of a batter to score runs over the six score labels. The players with lower values of entropy are mainly biased to one score level, $p_1$, compared to other score levels. In contrast, the players with higher entropy accumulate runs across all these six score levels, and thus, they can be described as balanced players. In other words, the entropy measures the diversity of a batter. Interestingly, the entropy attains a maximum value for Kohli; thus, it describes him as the most balanced or diverse player among these batters. The spreading of points about the linear regression line is minimal compared to other linear regression analyses discussed above. One can easily identify two clusters consistent with the proposed groups ``A" and ``B". 

The results discussed above assess the player's performance based on essential measures such as average runs, velocity, probability distribution vector, and entropy. However, these quantities fail to measure the similarities among the players accurately. To quantify similarities among the players, we have calculated the angle between two probability distribution vectors corresponding to each pair of players. Components of the probability distribution vectors are non-negative, giving rise to the value of the angles in between $0^\circ$ and $90^\circ$.
 The angle equal to zero describes that the corresponding players are identical according to their innings-wise runs distribution. On the other hand, the players are entirely non-identical if the angle between them is equal to $90^\circ $. Therefore, the lower value of the angle describes more similarities between the two players. 

\begin{figure}[H]
\centering
   \includegraphics[height=10cm, width=14cm,keepaspectratio]{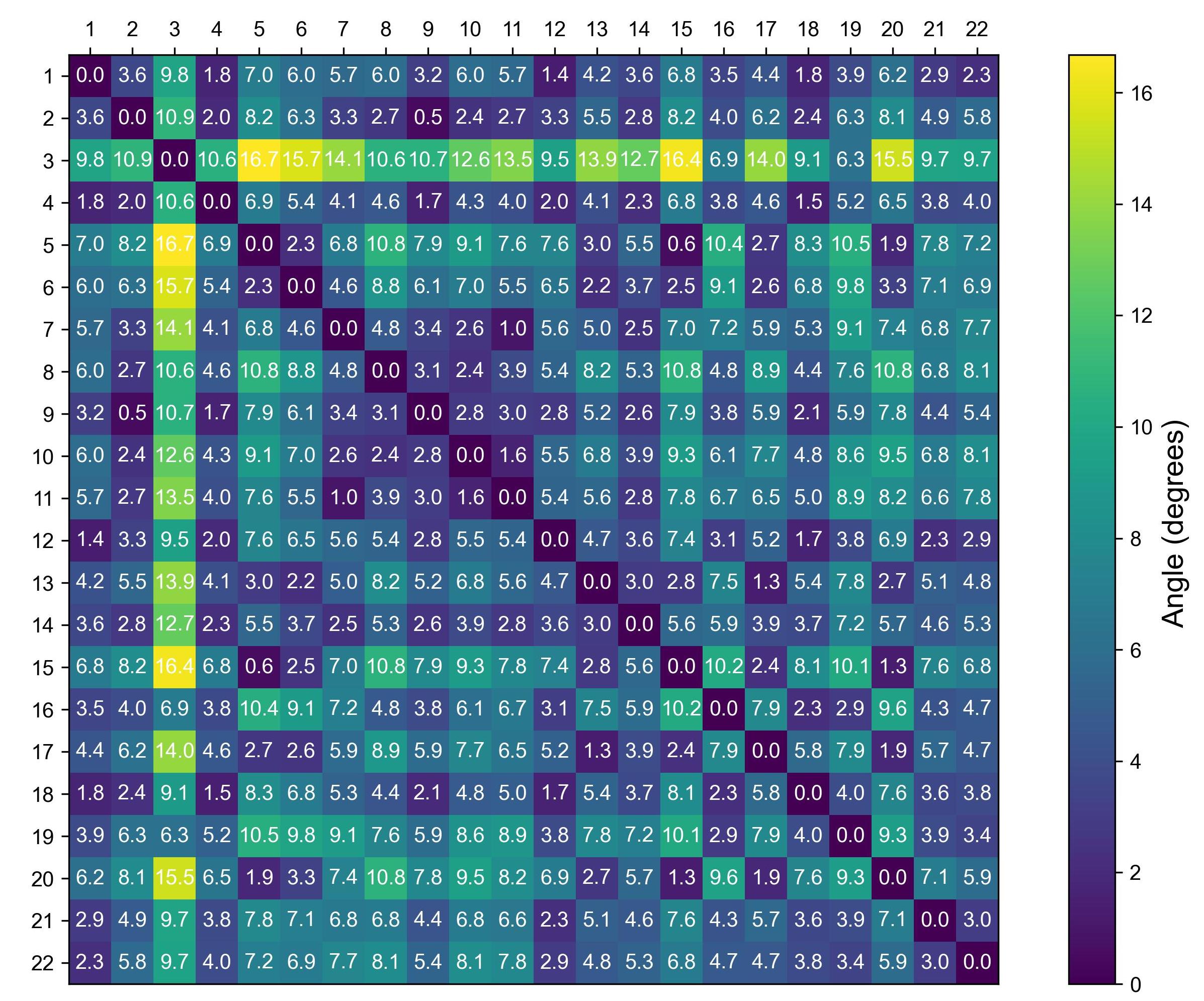}
   \caption{Color-coded representation of the angle matrix where the tick levels depicting serial numbers of batters. The color-coded bar represents a range of values for the angle between any two players. The lower value of the angle between two players gives rise to higher similarities. }
    \label{fig5}
\end{figure}

Figure~\ref{fig5} presents the angles between the probability distribution vectors of any two players in a matrix form.
 The minimum value of the angle is equal to $0.5^\circ$ for a pair of batters Sangakkara and Ganguly. 
According to the angle definition, these two players form the most identical pair among other pairs and possess similar distribution vectors. The angle attains a maximum value of $16.7^\circ$ for the players Virat and Jayasuriya. Thus, these two players possess fewer similarities than the other pairs. It should be noted that various small clusters can be found depending on the threshold values of the angle. For example, the value of angle less than equal to $2.0^\circ$ gives small clusters with similar types of players such as ( Jayasuriya, Dilshan, Gibbs ),  (Tendulkar, Ponting, Sharma ), ( Dravid, Dhoni ) and so on. The probability distribution vector corresponding to Kohli is different
from that of other players, making him a unique player. This is also reflected in the values of the angle between him and other players.

\begin{figure}[H]
\centering
   \includegraphics[clip=true,width=\columnwidth]{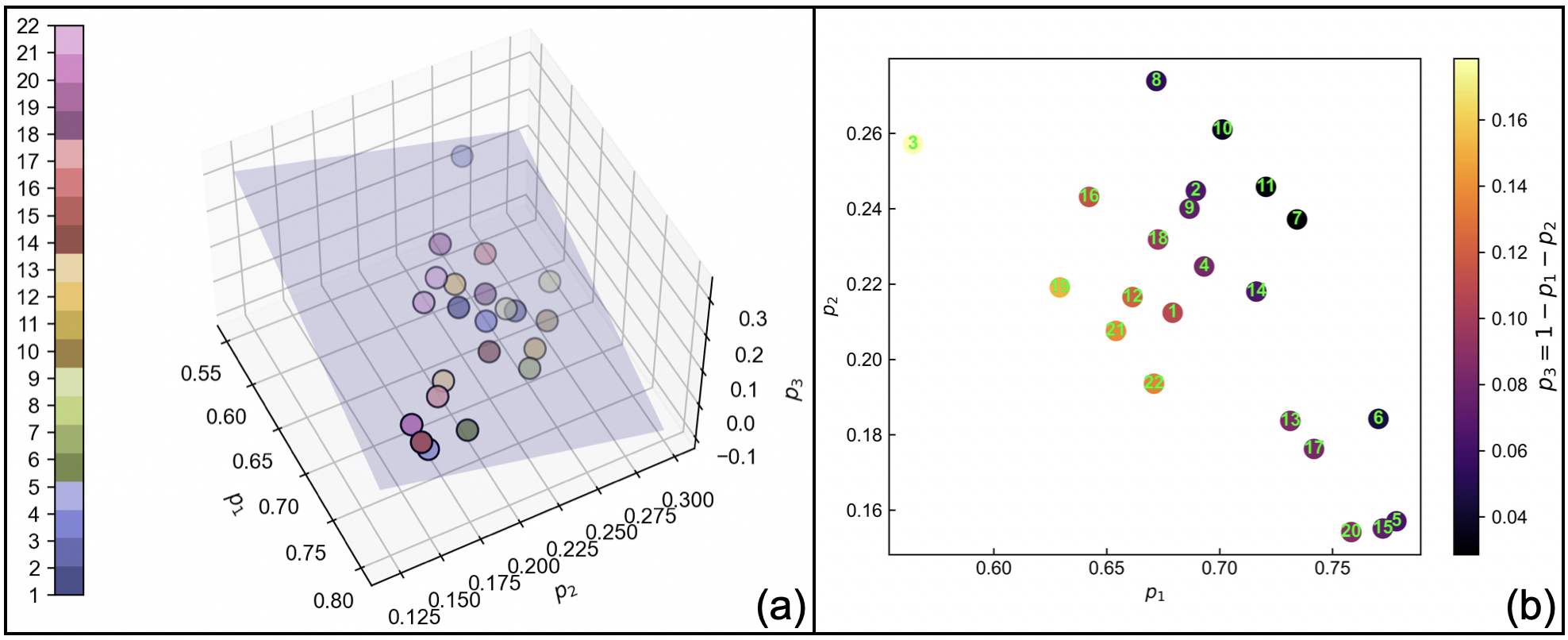}
   \caption{Three-dimensional representation of the positions of the batters in the probability space (a). All the points lying on the color-shaded plane are described by the equation~\ref{eq:closure_prob}. For convenience, (b) displaying the locations of all these points on the $p_1 -p_2$ projection plane. }
    \label{fig6}
\end{figure}
\begin{figure}[H]
\centering
   \includegraphics[clip=true,width=\columnwidth]{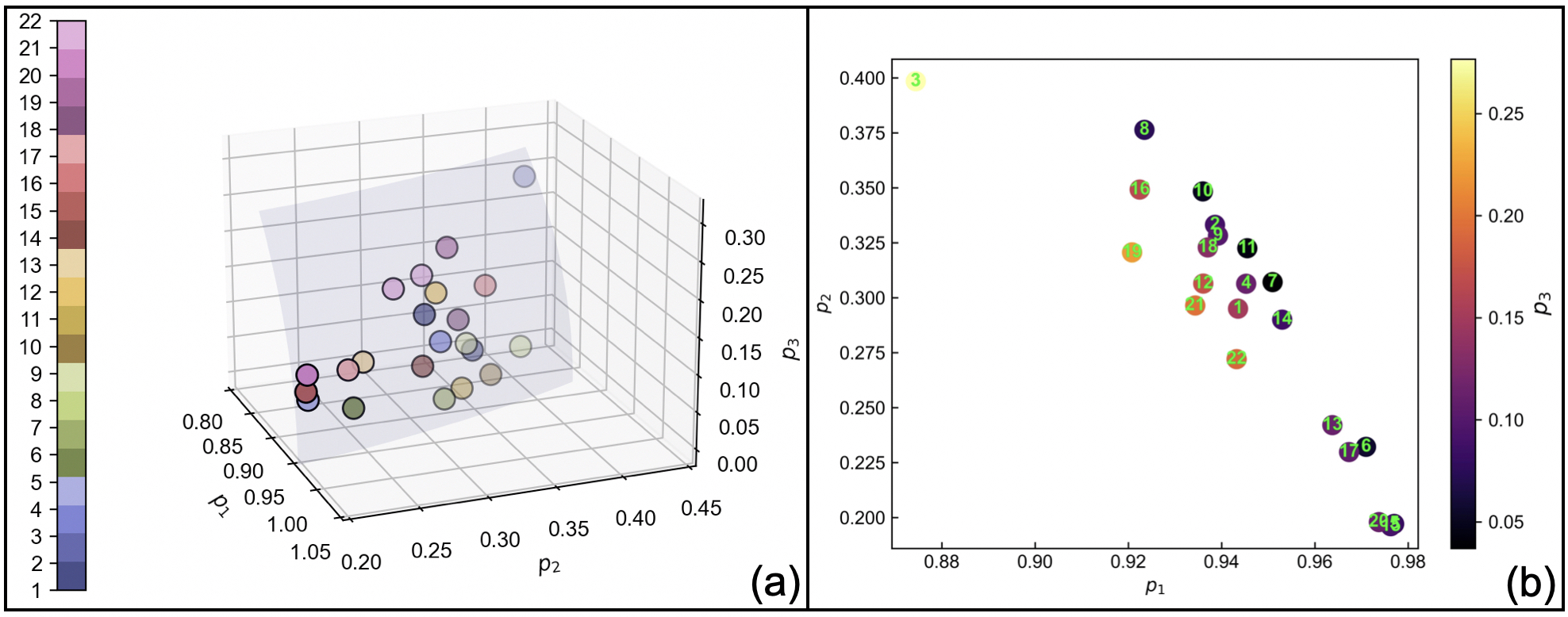}
   \caption{Three-dimensional representation of the positions of the batters in the probability space (a). All the points corresponding to the unit probability vectors are on the color-shaded cross-section of a spherical surface with a unit radius. For convenience, (b) depicting the location of all these points on the $p_1 -p_2$ projection plane. }
    \label{fig7}
\end{figure}

We have calculated the angle between any two players by considering six component unit vectors. However, the results discussed in the earlier sections describe that $p_1,p_2$, and $p_3$ are the main components of the probability distribution vectors, and other components are negligibly small compared to these three. Therefore, we have constructed three-dimensional probability distribution vectors to understand the angle matrix. The definition of $p_1$ and $p_2$ are remained unchanged.
Only, the probability $p_3$ is redefined as the probability of scoring run greater than equal to $100$. 
The complete closure relation of the probability distribution vectors is given by
\begin{equation}\label{eq:closure_prob}
 p_1+p_2+p_3=1.
\end{equation}
Therefore, all the points corresponding to the players lie on a plane, as shown in figure~\ref{fig6}(a). The components of the three-dimensional distribution vector are presented in figure~\ref{fig6}(b). It is found that $p_2$ decreases with increasing values of $p_1$. The component $p_3$ attains higher values for lower values of both $p_1$ and $p_2$ to ensure the complete closure relation~\ref{eq:closure_prob}. Furthermore, an empirical condition $p_1 \geq p_2 \geq p_3$ is found, which agrees with the earlier observation discussed above.
However, it is too hard to visualize the angle among the players from the components of distribution vectors as these are not unit vectors. We have transformed these distribution vectors into unit vectors. All the points corresponding to the players lie on the surface of a sphere of unit radius, as shown in figure~\ref{fig7}(a). Therefore, the corresponding unit vectors are the radial vectors of the sphere. Figure~\ref{fig7}(b) presents the components of these unit vectors. It should be noted that some data points overlap, perhaps giving rise to the similarities among the players. For example, data points corresponding to the batters Jayasuriya, Dilshan, and Gibbs superpose with each other, and their $p_3$ components are negligibly small. Therefore, the angles among these players are nearly equal to zero, indicating them to be similar types of batters. That is also reflected in the angle matrix data shown in figure.~\ref{fig5}. 
Other observations can be easily verified here by inspecting the angle matrix data.

Besides these studies discussed above, our approach can be utilized to assess the players in other formats of the game such as ``Test" and ``T20". The entropy can even be calculated for a batting-innings by counting runs ( single, double, triple, four, and six) that can assess the diversity of the innings and determine the all-time best innings performed by a team. Based on the similarity matrix study, one can classify identical players into groups, which can be utilized during player auctions. Furthermore, young/active players can compare their career trajectory with that of legendary players and improve their performance in the respective formats of the game. 
All of these studies deserve further investigation, and therefore, they are aimed for future research.

\section{Summary }
In summary, we have assessed twenty-two top batters selected on the basis of the total runs or centuries scored by the players in one-day cricket matches. It is found that the total runs of each player increases linearly with the number of innings. Also, the estimated runs rate obtained from the linear regression analysis increases linearly as a function of the average runs. In addition, we have calculated the probability distribution vector over six score levels. Some score levels are associated with the number of fifties and hundreds scored by the players. Two components of the probability distribution
vectors are found to vary linearly with average runs. Furthermore, we have computed the entropy for a player depending on the runs distribution over the six score levels. The entropy is maximum for Kohli, presenting him as the most balanced player among the batters. The entropy increases linearly with the average runs and classifies the players into two groups. Based on the entropy value, one can construct the player ranking chart, which directly measures the diversity of a player. In addition, the similarities among players are measured by computing the angle between the distribution vectors of two players. Some players are found to be almost identical to each other. 

\section{Acknowledgement}
We thank Sibaram Ruidas and Subham Ghosh for reading and providing critical comments to improve the manuscript.

\bibliographystyle{elsarticle-harv.bst}

\end{document}